\documentclass[onecolumn, showpacs, nofootinbib, aps]{revtex4}

\newcommand{\bi}{\bibitem}
\newcommand{\be}{\begin{eqnarray}}
\newcommand{\ee}{\end{eqnarray}}
\newcommand{\rar}{\rightarrow}

\begin{document}

\title{Reply to ``A note on the innocuous implications 
of a minimum length in quantum gravity'' by P.H.~Frampton}

\author{Cosimo~Bambi$^{\rm 1}$}
\email[E-mail: ]{cosimo.bambi@ipmu.jp}
\author{Katherine~Freese$^{\rm 2}$}
\email[E-mail: ]{ktfreese@umich.edu}

\affiliation{$^{\rm 1}$IPMU, University of Tokyo, 
Kashiwa, Chiba 277-8568, Japan \\
$^{\rm 2}$MCTP, University of Michigan, 
Ann Arbor, Michigan 48109, USA}

\date{\today}

\begin{abstract}
We reply to the comment ``A note on the innocuous implications 
of a minimum length in quantum gravity'' by P.H.~Frampton
[Class. Quantum Grav. 26 (2009) 018001] on our paper
``Dangerous implications of a minimum length in quantum gravity''
[Class. Quantum Grav. 25 (2008) 195013].
\end{abstract}


\maketitle

In ref.~\cite{noi}, we pointed out that the time--energy
Generalized Uncertainty Principle (GUP)~\cite{gup}
\be\label{eq-gup}
\Delta E \cdot \Delta t \gtrsim 1 + 
O \left(\frac{(\Delta E)^2}{M_{Pl}^2}\right)
\ee
seems to lead to a paradox when virtual Black Holes (BHs) are
taken into account: since larger and larger energy violations
would be less and less suppressed for $\Delta E > M_{Pl}$, 
super--Planck mass virtual BHs could induce dangerous processes 
in particle physics. In particular, we found the following
value for proton lifetime 
\be\label{eq-noi}
\tau_p \sim \frac{M_{Pl}^4}{m_p^5} \, \frac{1}{\eta^4} \, ,
\ee
when the decay is mediated by a virtual BH of mass $\eta M_{Pl}$,
with $\eta \gg 1$. In ref.~\cite{frampton}, the author suggested 
that actually one has to include an ``exponentially tunneling factor'' 
and there is no danger of conflict with experimental lower limit. 
His proton lifetime reads
\be\label{eq-frampton}
\tau_p \sim \frac{M_{Pl}^4}{m_p^5} \, \frac{1}{\eta^4} 
\, \exp \left[4\pi(\eta^2 - 1)\right] \, .
\ee
Here the term $\exp(-4\pi)$ is just a normalization factor. It 
was introduced in ref.~\cite{frampton} to have 
$\tau_p \sim M_{Pl}^4/m_p^5$ for $\eta = 1$.

First of all, we have to admit that it is difficult, or more
likely impossible, to say that our picture is correct: we
do not have any reliable theoretical framework in which we can 
perform the calculations, so all the considerations on the 
subject are necessarily based on more or less reasonable 
arguments, which can be questionable.

Bearing this in mind, we thoroughly agree with the author
of ref.~\cite{frampton} that {\it in general} there is such a
suppression factor, but we think it is inappropriate in the
context of the GUP. Classically, there is no time--energy
uncertainty relation and therefore tunneling processes are 
forbidden. In the standard picture of quantum mechanics, we 
can somehow say that energy can be violated for a short time. 
Larger and larger energy violations are more and more suppressed 
and we typically find the exponentially tunneling factor. 
In the case of the GUP, the situation is different, 
because eq.~(\ref{eq-gup}) says that we can have a huge violation 
of energy conservation for a very long time. It is thus natural 
to think that tunneling processes are still allowed, but that 
the exponential suppression factor has to disappear for 
$\Delta E > M_{Pl}$.

As discussed in ref.~\cite{frampton}, the exponential suppression
factor for heavy virtual BHs is already present in the
derivation of proton lifetime of ref.~\cite{kane}. The point is
that such a derivation is based on a path integral approach,
which ``knows'' the usual uncertainty principle, but does not 
the GUP: if we use standard technique, we implicitly assume 
standard physics and, in particular, the usual time--energy 
uncertainty relation. The path integral for gravity is
\be
Z \sim \int \mathcal{D}[g] \, 
\exp\left[\frac{i}{\hbar}S[g]\right] \, ,
\ee
where the integral is taken over all metric $g$.
Here we wrote explicitly $\hbar$ because it plays an
important role in our argument. The action for a single 
Schwarzschild BH of mass $\eta M_{Pl}$ in 4 dimensions is 
$S \sim \eta^2$. The path integral for $N$ non--interacting 
BHs evaluated in a box of volume $V$ turns out to be
\be\label{eq-path}
Z \sim \int_0^{\infty} d\eta \sum_{N=0}^{\infty}
\left(\frac{V}{L_{Pl}^3}\right)^N \frac{M_{Pl}}{N!}
\exp\left(-\frac{4\pi N \eta^2}{\hbar}\right) \, . 
\ee 
In the classical limit, $\hbar \rar 0$ and the contribution of
virtual BHs to any amplitude probability goes to zero, as it has 
to be expected. For $\hbar \neq 0$, we find the result of 
ref.~\cite{kane}: the spacetime is filled with Planck mass 
virtual BHs, the density is about one BH per Planck volume 
and the BH lifetime around one Planck time.

In the case of the GUP, we first notice that there are several
examples in the literature (see e.g.~\cite{examples}) which show
the GUP can be interpreted as a redefinition of $\hbar$
of the form
\be
\hbar \rar \hbar \left(1 + 
\alpha \frac{E^2}{M_{Pl}^2} \right) \quad {\rm or} \quad 
\hbar \left(1 + 
\alpha \frac{p^2}{M_{Pl}^2} \right) \, ,
\ee
where $\alpha$ is some positive constant of order one. We
would like to stress that this is a quite general result,
even if there is no proof that guarantees it is always true.
In the case of the amplitude probability of virtual BHs,
we would expect that the exponential factor in the path 
integral~(\ref{eq-path}) becomes
\be
\exp\left[-\frac{4\pi N \eta^2}{\hbar
\left(1 + \alpha \eta^2\right)}\right] \, . 
\ee
Proton decay can be thought as a process in which two
quarks of the proton are transformed into a quark/anti-quark 
and a charged lepton, via an intermediate virtual BH.
Instead of the formula~(\ref{eq-frampton}), we find the
following proton lifetime
\be\label{eq-final}
\tau_p \sim \frac{M_{Pl}^4}{m_p^5} \, \frac{1}{\eta^4} 
\, \exp \left[\frac{4\pi\eta^2}{1 + \alpha \eta^2} 
- \frac{4\pi}{1 + \alpha}\right] \, .
\ee
Since we need one virtual BH, we took the exponential factor 
with $N = 1$. The second term in the argument of the exponent, 
i.e. $4\pi/(1+\alpha)$, is only to have the same 
normalization of ref.~\cite{frampton}; that is, 
$\tau_p \sim M_{Pl}^4/m_p^5$ for $\eta = 1$. Eq.~(\ref{eq-final})
reduces to eq.~(\ref{eq-noi}) for $\eta \gg 1$ and 
confirms the idea of ref.~\cite{noi} that very heavy 
intermediate states are not suppressed.

\begin{acknowledgments}
C.B. was supported by 
World Premier International Research Center Initiative (WPI Initiative), 
MEXT, Japan. 
K.F. was supported in
part by the MCTP and DOE under grant DOE-FG02-95ER40899.  
\end{acknowledgments}

\end{document}